\documentclass[amsmath,aps,groupedaddress,showpacs,11pt]{revtex4-1} 

\usepackage{subcaption}

\usepackage[T1]{fontenc}

\usepackage{graphicx,xcolor,soul}
\usepackage{hyperref} 
\hypersetup{colorlinks=true,urlcolor=blue,citecolor=red,linkcolor=blue,filecolor=green}

\begin{document}

\title{Neutron-antineutron oscillations beyond the quasi-free limit}
\author{E. David Davis} 
\author{Albert R. Young}
\affiliation{Department of Physics, North Carolina State University, Raleigh, NC 27695-8202, USA,}
\affiliation{Triangle Universities Nuclear Laboratory, Durham, NC 27708-0308, USA}

\date{\today} 

\begin{abstract}
Prompted by plans to conduct a new neutron oscillation experiment at the European Spallation Source (ESS), 
we consider issues associated with the magnetic field that must be present, some of which are potentially exacerbated by
the significantly larger length $l$  contemplated for the neutron propagation region.
To this end, we introduce a stochastic model of the residual magnetic field within the propagation region 
which draws on features of magnetic profiles measured during the last free 
neutron oscillation experiment [conducted at the Institut Laue-Langevin (ILL)  in the 1990's]. We average over both fluctuations in the magnetic field 
sampled by neutrons, and 
representative spectra of neutron speeds. We find that \emph{deviations\/} from the quasi-free result for the antineutron probability do not 
depend quadratically on $l$ (as a naive perturbative estimate would suggest) but increase only linearly with $l$. As regards
the large spikes in the magnetic field which can be expected at, for example, joints in the magnetic shielding of the propagation region 
(despite compensating currents and magnetic idealization of the shield), we demonstrate that their effect scales as $l/D^{3/2}$, 
where $D$ is the diameter of the cylindrical magnetic shielding. Our arguments suggest that, provided the dimensions of the 
propagation region are such that the ratio $l/D^{3/2}$ does not exceed the value pertinent to the ILL experiment,
and these spikes occur close to either end of the propagation region, they can be neglected. We also establish that any large magnetic field 
encountered after the propagation region is exited will not diminish the probability for antineutron detection. For the range of values of $l$
of most interest to the ESS experiment, it should suffice to improve on the level of magnetic suppression achieved at the ILL by a factor of two.
\end{abstract}

\maketitle

\section{Introduction}\label{sc:Intro}

With the construction of the European Spallation Source (ESS)~\cite{ESS} in Lund, Sweden, there is renewed 
interest~\cite{Milstead2015,Theroine2015,Phillips2016,Frost2016} in an experimental study of free neutron-antineutron oscillations that 
would improve on the work done at the Institut Laue-Langevin (ILL) in the 1990's~\cite{Baldo-Ceolin1990,Baldo-Ceolin1994}, which, in turn,
superseded earlier experiments~\cite{Fidecaro1985,Bressi1989,Bressi1990}. As highlighted in the historical overview 
in Ref.~\cite{Mohapatra2014}, 
the search for neutron-antineutron oscillations would complement that for neutrino-less double 
beta decay. Both phenomena violate the ``accidental'' global anomaly-free Standard Model (SM) symmetry resulting in conservation of the
difference between the baryon number $B$ and the lepton number $L$. If the example of the SM has anything to teach us, it is that the 
identification of apposite symmetries is key to successful model building: thus, the status of this $(B-L)$-symmetry~\cite{Heeck2014,Addazi2016a} is 
important, and the \emph{observation\/} of neutron-antineutron oscillations [involving $|\Delta(B-L)|=2$ transitions] would be a discovery of 
physics beyond the Standard Model (or BSM physics). Furthermore, the detection of neutron-antineutron oscillations along with either 
a $B$- or a $(B-L)$-violating nucleon decay would imply~\cite{Babu2015a} that neutrino-less double beta decay must occur. 
In their own right, neutron-antineutron oscillation studies are an avenue to information on $|\Delta B|=2$ processes which can drive post-sphaleron 
baryogenesis~\cite{Babu2006,Babu2007,Babu2009,Babu2013,Patra2014}  and may 
contribute to an explanation of the observed baryon asymmetry of the universe. 
It has also been argued that, in view of their sensitivity to any difference in neutron and antineutron rest energies, neutron-antineutron oscillation 
experiments with free neutrons furnish a potentially stringent test of Lorentz invariance~\cite{Babu2015}  and, in similar vein, that discovery of 
neutron-antineutron oscillations would impose strong limits on any departure from the equivalence principle~\cite{Babu2016}. There have been formal
arguments~\cite{Gardner2015,Berezhiani2015,Fujikawa2015,Gardner2016} attempting to identify a role for \emph{CP\/} violation in neutron-antineutron 
oscillations, but the 
model-dependent considerations of Ref.~\cite{McKeen2016} suggest that it would be very small (although oscillations of 
heavy flavor baryons could exhibit substantially more \emph{CP\/} violation).

Models giving rise to neutron-antineutron oscillations date back (with the notable exception of Ref.~\cite{Kuzmin1970}) 
to the end of the 1970s, some of the pioneering papers being Refs.~\cite{Glashow1980,Mohapatra1980,Kuo1980,%
Chang1980,Deo1981,Nieves1981,Raychaudhuri1982,Fukugita1982,Lust1982,Ozer1982,Harari1982,Costa1982}. 
The advent of the Large Hadron Collider (LHC) has prompted an understandable emphasis on models with predictions which are 
testable at mass scales of $1~\text{TeV}/c^2$ or so (for the most up-to-date reviews, see Refs.~\cite{Phillips2016,Babu2013a} or, for a broader
perspective, Ref.~\cite{FileviezPerez2015}, which relies heavily on Refs.~\cite{Arnold2013,Fornal2014} in its discussion of 
neutron-antineutron oscillations). 
Despite the impending avalanche of LHC data on rare processes, a case is made in 
Ref.~\cite{Phillips2016} that the ESS-based effort to increase the lower bound on the neutron-antineutron oscillation rate will be useful.
This claim has been substantiated by a study~\cite{Calibbi2016} of simplified supersymmetric (SUSY) models with $R$-parity violation, 
which include $|\Delta B|=2$ processes and can accommodate the absence of SUSY signatures in run 1 of the LHC: even after allowing for
uncertainties~\cite{Martin2012} in hadronic matrix elements of $B$-violating interactions (which are 
being addressed by the lattice QCD community~\cite{Buchoff2012,Buchoff2016,Syritsyn2015}), it is concluded that the projected 
ESS experiment can probe for gluino and squark masses at energies beyond the foreseeable reach of the LHC program as well as other 
BSM searches based on flavor transitions, \emph{CP\/} violation and di-nucleon decays. Given the increasing number of 
models~\cite{Herrmann2014,Addazi2014,Brennan2015,Karozas2015,Addazi2016,Addazi2016b,Addazi2016c,Gu2016,Ramao2016}
which imply the possibility of observable neutron-antineutron oscillations, the more experiments performed the better.

In the mold of the ILL example, the current conception of the next generation of free oscillation experiment envisages a high flux of slow neutrons propagating 
down a long, horizontal, magnetically shielded, and evacuated cylinder (``the propagation region'') to a target  surrounded by an antineutron 
annihilation detector. Unique among experiments looking for $B$-violation, this setup offers the possibility of high sensitivity freed of 
significant backgrounds. The hope is that with a dedicated beamline, advances in neutron moderator technology, modern neutron optics, and 
a longer propagation region, the ILL limit on the free neutron oscillation probability can be bettered by at least three orders of magnitude.

In the latest assessment (conducted in Ref.~\cite{Phillips2016}) of future n$-\overline{\text{n}}$ oscillation experiments, 
it is suggested on the basis of a rough estimate that ``one requires a magnetic field in the $1-10\,\text{nT}$ regime to meet the
quasi-free condition'' and there is a call for ``a significant research program to understand how to achieve this lower limit 
[on the magnetic field] in a cost effective manner, and to understand the possible reduction in sensitivity that might arise from 
residual field configurations''. (The quasi-free condition is tantamount to the requirement that, for the passage through the propagation region, 
the difference in neutron and antineutron energies is much less than the limit set by the energy-time uncertainty principle.) 
The present paper represents a partial response to this challenge.

Our treatment of the residual magnetic field is based on an extension of the 4-state (or vector) model in 
Ref.~\cite{Kinkel1992} to accommodate random inhomogeneities. Inhomogeneities are perforce present when (mumetal) magnetic 
shielding is in place, because external magnetic fields, which are otherwise excluded, enter where segments of the shield are mechanically
joined together~\cite{Bitter1991,Dickerson2012}. Comparison of magnetic field profiles recorded during the ILL 
experiment~\cite{Dubbers2016}  points to the existence of unpredictable changes in inhomogeneities during runs and following the regular
magnetic idealizations~\cite{Bitter1991} of the shielding. By viewing 
the inhomogeneities as random, an interrelated effect can be incorporated, namely, the different values of the 
magnetic field experienced by neutrons on neighbouring flight paths through the propagation region. Differences in the magnetic field across a 
cross-section of the propagation region are due to inhomogeneities in the non-axial (or transverse) components of the field, which result 
from the increases in the longitudinal field at points along the shield where there are magnetic ``leaks''.
There are also random fluctuations with time in the ambient field although it has been past practice to compensate 
for these (whatever their source) by an active feedback system~\cite{Bitter1991}. An average over the ensemble of neutrons studied 
during the course of a many-year experiment involves implicitly an average over of the values of the actual field sampled.

A stochastic analysis of the effect of magnetic field on neutron-antineutron oscillations is clearly not needed if field profiles have been measured.
The measured profiles themselves can be used in conjunction with numerical solutions of Schr\"odinger's equation to determine the extent to which oscillations 
have been suppressed. However, our stochastic approach permits discussion of the design of an experimental set-up in the planning stage
for which there is, perforce, no actual magnetic profile data. 

In this paper, a number of topics related to generic aspects of the residual magnetic field are addressed. Paramount is the issue of whether, as suggested in 
Ref.~\cite{Phillips2016}, the acceptable size of the residual magnetic field in future experiments must be as low as  $1\,\text{nT}$ on average (i.e., about a 
factor of five smaller than the average field in the ILL experiment)? What, if any, is the cumulative impact of small unavoidable inhomogeneities in the field 
given the increased length of the propagation region to be used (maybe as much as nearly three times as long as that in the ILL experiment~\cite{Phillips2016})? 
The relation between the limit on the magnetic field and propagation distance has not been broached explicitly before but should be quantified. Another consequence of 
increasing the length of a mumetal shield which does not seem to have been given any attention (in the context of ILL-type experiments) is the increasing 
strength of spikes in the field~\cite{Dickerson2012} at magnetic leaks in the shielding. Even with carefully adjusted compensating currents 
(aimed at achieving cancellations accurate to more than one part in two thousand), large peaks ($\sim 0.1~\mu\text{T}$) usually persist at a handful of joints 
near the ends of the magnetic field profiles recorded during the ILL experiment (see, for example, Fig.~13 in Ref.~\cite{Bitter1991}). 
The dependence of these features on the length of the shield is non-linear. Thus, although the effect of these abrupt variations in the field was discounted in the 
analysis of the ILL experiment, their presence could be more significant for an experiment with a longer shield. There is also the large magnetic field in the region 
between the end of the mumetal shield and the annihilation detector to be considered.

In our stochastic model of the magnetic field, we distinguish between the large spikes in the field and the otherwise small residual field. The latter is treated 
perturbatively and its random behaviour (in the rest frame of any neutron) is assumed to be wide 
sense stationary~\cite{Yaglom1962},  in view of the presence of compensating currents designed both to iron out
inhomogeneities along the full length of the propagation region and actively counteract any time dependent fluctuations.
Our treatment of the large spikes is such that we do not have to commit ourselves to any detailed specification of their statistics.

The structure of the paper is as follows. Sec.~\ref{sc:Axial} begins with the 4-state model for neutron-antineutron oscillations as formulated in 
Ref.~\cite{Kinkel1992}, and then presents a perturbative result for the antineutron probability which includes the effect of small random 
inhomogeneities in the residual field (or magnetic noise).  In Sec.~\ref{sc:qfpe}, numerical estimates of the effect of this magnetic field noise 
on the quasi-free propagation efficiency $\eta$ for new ILL-type experiments are given, with particular attention being paid to dependence on the length 
of the shielded neutron propagation region. The impact of large fields at leaks in the shielding and at the end of the propagation region,
in the vicinity of the antineutron annihilation detector, is taken up in Secs.~\ref{sc:fjoints} and~\ref{sc:Dect}, respectively. Conclusions are drawn in Sec.~\ref{sc:Con}.

\section{$n-\overline{n}$ transitions in a small residual magnetic field \label{sc:Axial} }

For times relevant to ILL-type experiments (which are several orders of magnitude shorter than the lifetime of the neutron),
the Pauli-Schr\"odinger equation for the oscillating neutron-antineutron system in its rest frame reads~\cite{Kinkel1992}
 \begin{equation}\label{eq:ham2}
 i\hbar\frac{d\ }{dt} \begin{pmatrix}  \chi_\text{n}(t)  \\[0.25ex]  \chi_{ \overline{\text{n}}}(t)  \end{pmatrix} 
                  =  \hbar
                      \begin{pmatrix}
                           \frac{1}{2} \vec{\sigma}\cdot\vec{\omega}_L   & \delta\mathbf{1}_2                                            \\
                             \delta\mathbf{1}_2                                         & - \frac{1}{2} \vec{\sigma}\cdot\vec{\omega}_L
                      \end{pmatrix}\,
                     \begin{pmatrix}  \chi_\text{n}(t) \\[0.25ex]  \chi_{ \overline{\text{n}}}(t)   \end{pmatrix}  ,
\end{equation}
where $\chi_\text{n}$  ($\chi_{\overline{\text{n}}}$) are the neutron (antineutron) Pauli spinors, $\vec{\sigma}$ denotes the triplet of Pauli matrices 
$\{\sigma_x,\sigma_y,\sigma_z\}$, $\mathbf{1}_2$ is the $2\times 2$ unit matrix, 
$\delta$ is the matrix element of the scalar interaction coupling neutron and antineutron, and,
in terms of the residual magnetic field $\vec{B}$
(which is, in general, time-dependent), and the negative gyromagnetic ratio $\gamma$ of the neutron, the Larmor frequency vector
$\vec{\omega}_L(t)=-\gamma \vec{B}(t)$ (following the sign convention of Ref.~\cite{Levitt2008}).

In the limit that $\vec{B}$ is strictly uniform (and, hence, constant in the rest frame), Eq.~(\ref{eq:ham2}) can be solved exactly to yield
the Rabi-like formula for the probability of finding an antineutron a time $t$ after the oscillating state began as a neutron~\cite{Arndt1981}: 
\begin{equation}\label{eq:Rabi}
        P_{\overline{n}}  (t)     =    \frac{\delta^2}{\tfrac{1}{4}\omega_L^2+\delta^2}  \sin^2  \Big[ \big(\tfrac{1}{4}\omega_L^2+\delta^2 \big)^\frac{1}{2}  t  \Big] .
\end{equation}
Equation (\ref{eq:Rabi}) forms the basis for the identification of the \emph{quasi-free scaling\/} regime in which $\omega_L t\ll 1$: under this condition,
coupled with the existing empirical bound on $\delta$ (which implies that, in all cases of practical interest, $\delta\ll\omega_L$), it follows, from Eq.~(\ref{eq:Rabi}),
that  $P_{\overline{n}}(t)\approx \delta^2 t^2$, which is identical to the result expected in the absence of any magnetic field. 

In the rest frame of the oscillating system, inhomogeneities in $\vec{B}$ translate into explicit time-dependence, which, following~\cite{Kinkel1992},
can be accommodated by working with the interaction picture state vectors
\begin{equation}\label{eq:trtoI}
 \Psi_I (t) = \begin{pmatrix}
                   U_2(t) & 0 \\ 
                   0         & U_2^\dagger (t)
                  \end{pmatrix}                    \begin{pmatrix}
                                                           \chi_n\\[0.25ex]
                                                           \chi_{\overline{n}}
                                                          \end{pmatrix} ,
\end{equation}
where $U_2(t)$ describes evolution of the neutron spinor in the magnetic field $\vec{B}$, i.e.,
\begin{equation} \label{eq:eomU}
 i \frac{\partial\ }{\partial t} U_2(t) = \tfrac{1}{2} \vec{\sigma}\cdot\vec{\omega}_L \, U_2(t) 
\end{equation}
with $U_2(0)=\mathbf{1}_2$. These interaction picture state vectors satisfy the equation of motion
\begin{displaymath}
 i\hbar \frac{d\Psi_I }{dt}  =  \hbar\delta   \begin{pmatrix}             0                   &   \big( U_2^\dagger(t) \big)^2    \\[0.5pc]
                                                                                       \big( U_2(t) \big)^2     &                     0                            \end{pmatrix}   \Psi_I  .
\end{displaymath}
To lowest order in $\delta$, the corresponding probability for detection of an antineutron (with a polarization insensitive detector), starting from a source of 
unpolarised neutrons at time $t=0$, is
\begin{equation}\label{eq:pvv}
 P_{\overline{n}} (t) = \frac{\delta^2}{2}  \int\limits_0^t dt^\prime\! \int\limits_0^t dt^{\prime\prime}\,
                                                            \mathrm{Tr} 
                                                            \left[ \big( U_2^\dagger (t^\prime)\big)^2 \big( U_2 (t^{\prime\prime}) \big)^2 \right]  ,
\end{equation}
a result first obtained in Ref.~\cite{Kinkel1992} (where the matrix $U_2$ is denoted by $\Phi$).

Equation (\ref{eq:pvv}) permits, in principle, a non-perturbative treatment of the magnetic field, but, in the estimates of the quasi-free
propagation efficiency of interest, a perturbative calculation suffices. For the purpose of calculating the trace in Eq.~(\ref{eq:pvv}) to quadratic 
order in small magnetic fields (equivalently, small $\vec{\omega}_L$), $U_2(t)$ can be approximated as $\exp[-\tfrac{i}{2} \vec{\varphi}(t)\cdot\vec{\sigma}]$,
where the dynamical phase vector
\begin{equation}
 \vec{\varphi}(t) \equiv \int\limits_0^t dt^\prime \vec{\omega}_L(t^\prime) .
\end{equation}
To this order in $\omega_L$, Eq.~(\ref{eq:pvv}) reduces to
\begin{equation}\label{eq:bs}
 \frac{P_{\overline{n}}(t)}{\delta^2\, t^2}  =  1 - 
           \frac{1}{t} \int\limits_0^t dt^\prime \Bigg| \vec{\varphi}(t^\prime )  - 
                              \frac{1}{t} \int\limits_0^t dt^{\prime\prime} \vec{\varphi}(t^{\prime\prime})   \Bigg|^2 + \ldots ,
\end{equation}
which is the same as Eq.~(18) in Ref.~\cite{Kinkel1992}. 

Allowance can be made for the randomness of the values of magnetic field inhomogeneities sampled in the propagation region by viewing the antineutron 
detection probability $P_{\overline{n}}$ in Eq.~(\ref{eq:pvv}) as a functional of a random (or stochastic) process~\cite{Gardiner2009}, namely, $\vec{\omega}_L(t)
=-\gamma \vec{B}(t)$, with the probability distribution functional ${\mathcal  P}[ \vec{\omega}_L(\cdot) ]$. (In what follows, a pair of angle 
brackets $\langle\ldots\rangle$ will denote an expectation value computed with the probability distribution functional ${\mathcal  P}[ \vec{\omega}_L(\cdot) ]$.)
In line with the discussion in the introduction, it will be assumed that the process 
$\vec{\omega}_L(t)$ is wide-sense stationary, i.e., the expectation value $\langle \vec{\omega}_L\rangle$ is time-independent and the 
auto-correlation functions $ \left\langle\, \bigl( \omega_{L,i}(t_1) - \langle \omega_{L,i} \rangle \bigr) \bigl(\omega_{L,j}(t_2) - \langle \omega_{L,j} 
\rangle \bigr)\, \right\rangle $ involving Cartesian components $\omega_{L,i}$ depend only on the relative time difference $t_1-t_2$.

Under this plausible assumption about the statistics of $\vec{\omega}_L$, the expectation  value of the probability in Eq.~(\ref{eq:bs}) can be expressed in
terms of the sum of the power spectral densities
\begin{equation}\label{eq:PSDdef}
  S_{\omega_L}^{(i)} (\omega ) =  \int\limits_{-\infty}^{+\infty} \left\langle\, \bigl(\omega_{L,i}(\tau) - \langle \omega_{L,i} \rangle \bigr)
                                                                                                   \bigl(\omega_{L,i}(0) - \langle \omega_{L,i} \rangle \bigr)\, \right\rangle  e^{-i\omega\tau} d\tau .
\end{equation}
Thus, for example,
\begin{displaymath}
 \Big\langle \frac{1}{t} \int\limits_0^t dt^\prime \vec{\varphi}(t^\prime )\cdot\vec{\varphi}(t^\prime ) \Big\rangle 
          = \tfrac{1}{3} \big| \langle \vec{\omega}_L \rangle\big|^2 t^2 
                 +  \frac{1}{\pi} \int\limits_{-\infty}^{+\infty}  \frac{d\omega}{\omega^2}  \big[ 1 - \text{sinc} (\omega t) \big]  S_{\omega_L} (\omega ) ,
\end{displaymath}
where $S_{\omega_L} (\omega )= \sum_i S_{\omega_L}^{(i)} (\omega ) $
and $\text{sinc}(x)$ denotes the unnormalised cardinal sine function [i.e., $\text{sinc}(x)\equiv\sin (x)/x$ for $x\not=0$ with $\text{sinc}(0)=1$]. 
The complete result for $\langle P_{\overline{n}}(t) \rangle$ reads
\begin{equation}\label{eq:wkFld}
\frac{  \langle P_{\overline{n}}(t) \rangle }{\delta^2 t^2} =  1 -  \tfrac{1}{12} \big| \langle \vec{\omega}_L \rangle \big|^2 t^2
                                                       - \frac{1}{2\pi} \int\limits_{-\infty}^\infty \frac{d\omega}{\omega^2} \left[1 - \text{sinc}^2 (\tfrac{1}{2}\omega t  )  \right] 
                                                        S_{\omega_L} (\omega ) + \ldots\, .
\end{equation}
An immediate implication of Eq.~(\ref{eq:wkFld}) is that the deviation of $\langle P_{\overline{n}}(t) \rangle$ from the quasi-free estimate (of $\delta^2 t^2$)
increases with increasing time-of-flight $t$.

For the purpose of making numerical estimates, the power spectral densities $S_{\omega_L}^{(i)} (\omega )$ will be taken to be the Lorentzians
\begin{equation}\label{eq:mwssn}
 \frac{2\lambda_i}{1+\tau_c^2\omega^2}  ,
\end{equation}
defining wide-sense stationary Markovian noise. Linear superpositions of such Markovian noise sources can model ubiquitous $1/f^\alpha$ noise 
($0<\alpha<2$)~\cite{Kuopanportti2008,Gorman2012}. In Eq.~(\ref{eq:mwssn}), $\tau_c$ is a correlation time, arising from spatial 
correlations in the longitudinal (or axial) direction of the propagation region and proportional to the associated correlation length
$l_c$ (defined in the laboratory frame). Two choices of $l_c$, bracketing the range of reasonable values, are adopted in Sec.~\ref{sc:qfpe}: $l_c=10\,\text{m}$ 
(about twice the distance between adjacent joints in the shielding used in the ILL experiment) and $l_c=0$ (the ``white'' noise limit). 
The strength $\lambda_i$ can be related to the statistics of the residual magnetic field 
by considering the expectation value (with respect to ${\mathcal  P}[ \vec{\omega}_L(\cdot) ]$) of the \emph{square\/} of
\begin{displaymath}
          \Delta \omega_{L,i} = \frac{1}{t} \int\limits_0^t dt^\prime \big( \omega_{L,i}(t^\prime ) - \langle \omega_{L,i}  \rangle \big) .
\end{displaymath}
Computation of $\sigma_{L,i}^2\equiv\langle \big(\Delta \omega_{L,i}\big)^2 \rangle$ with Eq.~(\ref{eq:mwssn}) (under the assumption of wide-sense stationarity)
implies that 
\begin{equation}\label{eq:lam}
  \lambda_i =\tfrac{1}{2}\sigma_{L,i}^2\, t/\beta(\tfrac{t}{\tau_c}) 
\end{equation}
with $\beta(x)\equiv 1-(1-e^{-x})/x$.
Equation (\ref{eq:wkFld}) depends on the
combination $\sum\limits_i \lambda_i$ or $\sigma_{L}^2\equiv\sum\limits_i \sigma_{L,i}^2$.

In principle, $\langle \omega_{L,i}\rangle$ and $\sigma_{L,i}$ are to be extracted from the detailed comparison of magnetic field profiles
along the length of the propagation region. 
In as much as inhomogeneities in a given profile and changes from 
one profile to the next both result from the influence of magnetic leaks, it is reasonable to
suppose that \emph{estimates\/} of $\langle \omega_{L,i}\rangle$ and $\sigma_{L,i}$ can be inferred from a \emph{single\/} profile (of the $i$th
component): $\langle \omega_{L,i}\rangle/\gamma$ and $\sigma_{L,i}/|\gamma|$ should be comparable to the mean and standard deviation, 
respectively, of the spatial variations in this profile. This assertion can be bolstered by an appeal to the ergodic properties expected of 
wide sense stationary random processes (see~Sec.~4 in Ref.~\cite{Yaglom1962}), which imply that the averages over the 
ensemble of realisations of $\omega_{L,i}$ defining its expectation value and auto-correlation function can be replaced by averages of typical
realisation $\omega_{L,i}^{(*)}$ over an infinite time. (Temporal averages in the rest frame become spatial averages in the laboratory frame.)

An eyeballing of field profiles~\cite{Dubbers2016} from the ILL experiment (including Fig.~13 in Ref.~\cite{Bitter1991})
suggests that, except in the vicinity of a few isolated magnetic leaks (considered in Sec.~\ref{sc:fjoints}),  the mean of the residual 
\emph{axial\/} field component $B_\|$ and its standard deviation (about this mean) were reduced to the level  of $5\,\text{nT}$ or so.
There is less information on the transverse components (as the single available profile is corrupted by a magnetised screw), but,
because of the coupling between inhomogeneities of axial and transverse components, it can be assumed that their means and standard deviations are similar.
In fact, in Ref.~\cite{Bitter1991}, it is claimed that the transverse components are smoother than the axial component
and that their collective effect is at most equal to that of the axial component --- see the discussion immediately preceding Eq.~(12) in 
Ref.~\cite{Bitter1991}.  In Sec.~\ref{sc:qfpe}, we set
\begin{equation}\label{eq:magpar}
\big| \langle \vec{\omega}_L \rangle\big|^2 = 2\gamma^2(5\,\text{nT})^2 = \sigma_{L}^2 ,
\end{equation}
corresponding to the conservative selection of a mean of $5\,\text{nT}$ and a the standard deviation of  $5\,\text{nT}$ for \emph{both\/} the axial component $B_\|$ and
the \emph{total\/} transverse component $B_\perp$ of the residual magnetic field $\vec{B}\,[=\vec{B}_\|+\vec{B}_\perp]$. 

 \section{Estimates of the quasi-free propagation efficiency\label{sc:qfpe}}

A figure of merit for the prospects of an ILL-type experiment is proportional to the product $\eta \cdot \langle t^2 \rangle_v$, where 
the subscripted brackets $\langle \ldots\rangle_v$ denote an average over the neutron time-of-flight spectrum and the quasi-free propagation efficiency
\begin{equation}\label{eq:Bseta}
 \eta \equiv  \frac{ \langle \langle P_{\overline{n}}(t) \rangle \rangle _v}{\delta^2  \langle t^2 \rangle_v} .
\end{equation}
Existing studies~\cite{Phillips2016} of the optimal length $l$ of the shielded propagation region for an ILL-type experiment at the ESS
have taken into account the increase in $\langle t^2 \rangle_v$ associated with an increase in $l$. 
We now investigate the the impact on $\eta$ of an increase in $l$.

Neutron optics involving elliptical super-mirrors which will redirect neutrons to the annihilation target is
crucial to plans for future ILL-type experiments, but, to begin with, it will be supposed that neutrons propagate
directly from the source to the target through a fixed horizontal distance $l$. It will also be assumed that the 
entirety of this distance is magnetically shielded. (This last assumption is justified in Sec.~\ref{sc:Dect}.)
 
Under these assumptions, $\langle P_{\overline{n}}(t)\rangle$ can be re-interpreted as the probability of detecting
an antineutron with an axial component $v$ of velocity equal to $l/t$. 
One can define an antineutron detection probability which is a function of $v$: $ P_{\overline{n},\text{v}}(v)\equiv
\langle P_{\overline{n}}(t=l/v) \rangle$.
 The average $\langle \langle P_{\overline{n}}(t) \rangle\rangle_v$ in Eq.~(\ref{eq:Bseta}) is found by integrating 
$P_{\overline{n},\text{v}}(v)$ over the spectrum of axial speeds. 

The total antineutron probability for a propagation region of length $l$ is
\begin{equation}
 P_{\overline{n},l}     =    \left\langle P_{\overline{n},\text{v}}(v) \right\rangle_v 
                             \equiv \left.  \int\limits_{v_\text{min}}^\infty P_{\overline{n},\text{v}}(v) n(v) dv \right/   \int\limits_{v_\text{min}}^\infty n(v) dv ,
\end{equation}
where $n(v)$ is the \emph{probability density\/} for the axial speed $v$ during the experiment
and the positive lower limit $v_\text{min}$ is the smallest axial speed which is consistent with the requirement that the oscillating neutron-antineutron system,
 which is in free fall, traverse the \emph{horizontal\/} propagation region without hitting its tubular walls.
(Such collisions have to avoided for the same reason that the propagation region must be evacuated.) In the quasi-free limit, $P_{\overline{n},l}$ reduces to  
$P_{\overline{n},l}^0 = \delta^2 l^2  \left\langle v^{-2} \right\rangle_v$. Thus, the quasi-free propagation efficiency is
\begin{equation}\label{eq:etav}
   \eta \equiv  P_{\overline{n},l} / P_{\overline{n},l}^0
                  =   \left. \int\limits_{v_\text{min}}^\infty \frac{ P_{\overline{n},\text{v}}(v)}{\delta^2l^2} n(v) dv \right/ \int\limits_{v_\text{min}}^\infty v^{-2} n(v) dv ,
\end{equation}
which, on substitution of the expression for $P_{\overline{n},\text{v}}(v)~[=
\langle P_{\overline{n}}(t=l/v) \rangle]$ implied by Eqs.~(\ref{eq:wkFld}), (\ref{eq:mwssn}) and (\ref{eq:lam}) 
(with replacement of $t$ and $\tau_c$ by $l/v$ and $l_c/v$, respectively),
becomes
\begin{equation}\label{eq:etafinal}
 \eta = 1 -  \tfrac{1}{12}\Big[  \big| \langle \vec{\omega}_L \rangle \big|^2
                                      +        2\widetilde{\beta}(\tfrac{l_c}{l}) \sigma_L^2    \Big] \frac{ \mu_{(-4)} }{ \mu_{(-2)} }\,   l^2 + \ldots ,
\end{equation}
where 
\begin{displaymath}
 \mu_{(k)} \equiv  \int\limits_{v_\text{min}}^\infty v^k n(v) dv 
\end{displaymath}
and $\widetilde{\beta}(x)\equiv[1-3x+6x^2\beta(1/x)]/\beta(1/x)$. 

The apparent quadratic dependence of $\eta$ on $l$ is modified by the factor $\widetilde{\beta}(l_c/l)$ and the ratio $\mu_{(-4)}/\mu_{(-2)}$ through its 
dependence on $v_\text{min}$
(see the next paragraph). The function $\widetilde{\beta}(x)$, which regulates the contribution of the fluctuation term (i.e., the term containing $\sigma_L^2$), is unity 
for $l_c=0$ and
decreases monotonically as $x=l_c/l$ increases: as one should expect, increasing correlations diminish the effect of fluctuations.

An elementary estimate of the extent free fall (ignoring any influence of the neutron optics) suggests that $v_\text{min}$ should be approximately proportional 
to $l$, with a constant of proportionality $\alpha\approx(\tfrac{1}{2}g/d)^{1/2}$, where $g$ is the acceleration due to gravity, and $d$ is the vertical
distance through which the oscillating neutron-antineutron system can fall and yet still strike the detector (without interacting with the confining walls of the 
propagation region). Values of $d$ consistent with current plans (as in Fig.~3 of Ref.~\cite{Phillips2016}) for future experimental setups 
range from about 1~m ($\alpha\approx 2.2\,\text{s}^{-1}$) to about 2~m ($\alpha\approx 1.6\,\text{s}^{-1}$). Accordingly, in the present investigations,
we set $v_\text{min}=(1.9\,\text{s}^{-1}) l$.

Two quite different choices of $n(v)$ are made. One is the physically motivated superposition 
\begin{equation}\label{eq:ns}
n_s(v) =  (1-f) \frac{4}{\sqrt{\pi}} \frac{v^2}{v_T^3} \exp\left(-\frac{v^2}{v_T^2} \right)   + f\,  \frac{v_c}{v^2} \Theta(v-v_c)  ,
\end{equation}
where $f$ is the epithermal fraction, $v_T$ is the most probable speed for the Maxwellian component and
$v_c$ is the epithermal cutoff speed, parametrised in terms of $v_T$ as $v_c=\sqrt{\mu}\,v_T$~\cite{Gould2006}. 
(In terms of the absolute temperature $T$ of the Maxwellian, Boltzmann's constant $k$ and the neutron mass $m$, $v_T=\sqrt{2kT/m}$.)
The other more \emph{ad hoc\/} selection is the Nakagami-like probability density function
\begin{equation}\label{eq:nd}
n_d(v) = \frac{2\Omega^{-\nu}}{\Gamma(\nu)}(v-v_\text{min})^{2\nu-1} \exp\left[-\frac{1}{\Omega}(v-v_\text{min})^2\right] \Theta(v-v_\text{min}) ,
\end{equation}
where $\Gamma(\nu)$ is the gamma function, and $\Omega=(v_\text{max}-v_\text{min})^2/(\nu-\tfrac{1}{2})$, which ensures that $n_d(v)$ 
peaks at $v=v_\text{max}$. For $T=35\,\mbox{K}$ (typical of cold neutrons), $\mu=5$ (advocated in Ref.~\cite{Gould2006}) and $f=0.15$ 
(results are insensitive to an increase or decrease in $f$ by a factor of 3), $n_s(v)$ bears a reasonable resemblance to the simulated spectrum of speeds
for the cold neutron source considered in Ref.~\cite{Phillips2016}, while, if $v_\text{max}=800\,\text{m/s}$ and $\nu=\tfrac{3}{2}$,  $n_d(v)$ is 
similar to unpublished simulation results~\cite{Snow2014} for the speeds of neutrons reaching the detector with neutron optics of the kind outlined in
Ref.~\cite{Phillips2016}. Use of both $n_s(v)$ and $n_d(v)$ allows one to gauge the uncertainties in estimates of $\eta$ due to uncertainties in 
$n(v)$. 

{
\begin{figure}[t]\centering
\includegraphics[height=6cm]{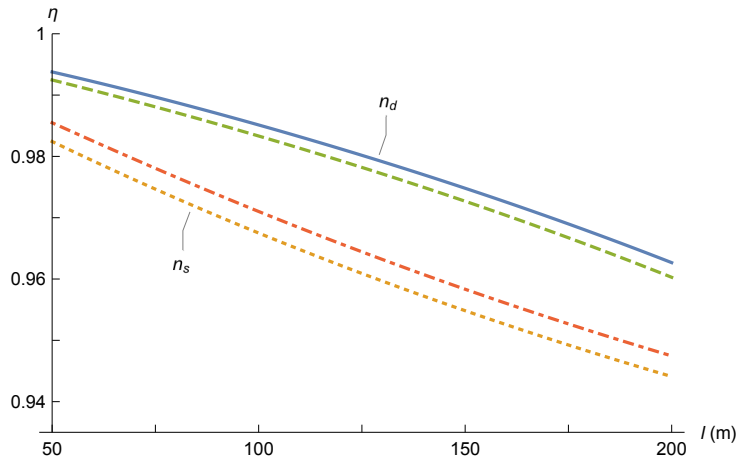}
\caption{The quasi-free propagation efficiency $\eta$ versus length $l$ of the propagation region  
when the mean \emph{and} the standard deviation of the axial component and the \emph{total\/} transverse component of the residual
magnetic field are both equal to $5\,\text{nT}$.
The upper (lower) pair of curves corresponds to the harder (softer) spectrum $n_d$ ($n_s$) of axial speeds.
Within each of these pairs of curves, $l_c=10\,\text{m}$ ($l_c=0$) for the upper (lower) curve.
 (Parameters of $n_s$ and $n_d$ are given in the text.)  \label{fg:etaL}
} 
\end{figure}
}   

Figure~\ref{fg:etaL} contains plots of the quasi-free propagation efficiency $\eta$ versus shield length $l$  for the four possible pairings of noise model 
($l_c=0$ versus $l_c=10\,\text{m}$) with speed spectrum ($n_s$ or $n_d$).  
Both the axial and non-axial components, $\vec{B}_\|$ and $\vec{B}_\perp$, respectively, of the magnetic field $\vec{B}$ have a 
mean and standard deviation of $5\,\text{nT}$ [see~Eq.~(\ref{eq:magpar})].

With regard to the proposed ILL-type experiment at the ESS, the deliberations in Ref.~\cite{Phillips2016} have not ruled out the possibility that
the magnetic field in the propagation region may need to be suppressed to as low as $1\,\text{nT}$ . However, it would seem from Fig.~\ref{fg:etaL} 
that a larger residual magnetic field comparable to that achieved in the ILL experiment can be tolerated: $\eta$ is still in excess of 0.94 for the most 
interesting values of $l$, identified as between $175\,\text{m}$ and $200\,\text{m}$ in Ref.~\cite{Phillips2016}. Although, as anticipated, $\eta$ 
decreases with increasing $l$, the extent of this decrease is sufficiently small ($< 3\%$ as $l$ increases from $100\,\text{m}$ to $200\,\text{m}$) that 
it can be disregarded in any determination of the optimal length of the propagation region.

In view of the quadratic dependence of $\eta$ in Eq.~(\ref{eq:etafinal}) on magnetic field strength, an improvement by a factor of two on the level
of field suppression attained in the ILL experiment 
will be enough to guarantee quasi-free propagation efficiencies of more than $98\%$.
Furthermore, given its close proximity to unity, $\eta$ itself can then be ignored for the purposes of  calculating a figure of merit for an ILL-type experiment, as, indeed, it 
was in Ref.~\cite{Phillips2016}.

It is apparent from Fig.~\ref{fg:etaL} that the nature of the axial speed spectrum $n(v)$ has a bigger impact on $\eta$ than the character of the noise in the residual 
magnetic field. Another inference from Fig.~\ref{fg:etaL} is that, in lieu of adequate empirical information on either $n(v)$ or the character of the magnetic field noise, 
one can rely on the estimate of $\eta$ with
white noise ($l_c=0$) and the Maxwellian-plus-epithermal spectrum $n_s(v)$ to be the most conservative.
 
The effect of neutron optics has been omitted in the above calculations. The presence of elements like focussing reflectors  would mean that there is a range of 
flight paths for a given time of flight $t$. However, given the weak and approximately \emph{linear\/} character of the dependence  of 
$\eta$ on $l$, the impact of the dispersion in flight paths on the propagation efficiencies in Fig.~\ref{fg:etaL} should be negligible except for the fact that the abscissa
$l$ should be reinterpreted as the \emph{average\/} flight path.
    
\section{Effect of localised large fields at magnetic leaks}\label{sc:fjoints}

In the previous section, the large fields in the immediate vicinity of magnetic leaks in the mumetal shielding were ignored.
We now present an argument that justifies neglect of these large fields. It rests largely on the fact that they are localised or, more graphically, ``spike-like''. 

It is convenient to begin with a one-dimensional treatment which takes into account only the axial component of magnetic 
fields. In this case, the exact solution of Eq.~(\ref{eq:eomU}) can be obtained in closed form as
$U_2(t) = \exp[  -\tfrac{i}{2}  \varphi_z(t)\sigma_z ]$, 
where the axial direction within the propagation region has been identified with the $z$-axis (by assumption, $U_2(0)$ is the 
identity matrix $\mathbf{1}_2$), and Eq.~(\ref{eq:pvv}) for the probability of an antineutron reduces to~\cite{Kinkel1992} 
\begin{equation}\label{eq:pvv1D}
 P_{\overline{n}} (t) =  \delta^2 \left|  \int\limits_0^t\!  dt^\prime  \exp\big[i\varphi_z(t^\prime )\big]   \right|^2 .
\end{equation}
The breakout of (axial) magnetic field at points along the shielding can be modelled by replacing the random process $\omega_{L,z}(t)$
by the sum
\begin{equation}\label{eq:delang}
   \omega_{L,z}(t) + \sum\limits_{k=1}^N \Delta\varphi_z^k\, \delta(t-t_k^*) ,
\end{equation}
where $\Delta\varphi_z^k$ denotes the net change in the $z$-component of the dynamical phase across the $k$th leak (which is 
encountered at time $t_k^*$ or, in the laboratory frame, at an axial distance $l_k^*$ from the beginning of the propagation region).
For neutrons of a given axial component $v$ of velocity, the $\Delta\varphi_z^k$'s can be assumed to be constant during runs of the 
experiment between any two successive magnetic idealizations of the shield.
(Only the process~\cite{Bitter1991} whereby the magnetic shield is idealized is likely to give rise to substantial changes in the $\Delta\varphi_z^k$'s.)

The issue of whether the $\Delta\varphi_z^k$'s in Eq.~(\ref{eq:delang}) induce significant corrections to the estimate of $\eta$ in Eq.~(\ref{eq:etafinal})
can be addressed by considering the maximum value $\Delta\varphi_\text{max}$ of the $|\Delta\varphi_z^k|$'s. 
If the largest spike in the axial field attains a value of magnitude $B_\text{max}$ and has a half-width of $\Delta l$, then $\Delta\varphi_\text{max}$ may be 
calculated as 
\begin{displaymath}
\Delta\varphi_\text{max} \approx |\gamma| B_\text{max} \frac{\Delta l}{v_\text{min}}  \approx \Delta\varphi_\text{max}^\text{ILL} \left[\frac{v_\text{min}}{B_\text{max}}\right]_\text{\!ILL}
                                                                                                                                             \frac{B_\text{max}}{v_\text{min}} 
                                                                                                          \approx \Delta\varphi_\text{max}^\text{ILL}\, \left(\frac{D_\text{ILL}}{D}\right)^{3/2}\, \frac{l}{l_\text{ILL}} .
\end{displaymath}
In rewriting $\Delta\varphi_\text{max}$ in terms of $\Delta\varphi_\text{max}^\text{ILL}$, it has been assumed that $\Delta l$ is not significantly affected by changes 
in the length $l$ of the shielding, and that, over the relevant range of $l$, $B_\text{max}$ is approximately 
proportional to $(l/D)^2$ \cite{Dickerson2012,Mager1970,Chen2006,Sun2013}, where $D$ is the diameter of the shielding, while 
$v_\text{min}$ is taken to be approximately proportional to $l/D^{1/2}$ (as in the free-fall based estimate of $v_\text{min}$ earlier). 

As regards the configurations for future ILL-type experiments discussed in Ref.~\cite{Phillips2016}, even in the worst case contemplated of smallest $D$ and largest $l$  
($D=2\,\text{m}$, $l=200\,\text{m}$), $\Delta\varphi_\text{max} \approx 0.9\, \Delta\varphi_\text{max}^\text{ILL}$. Provided \emph{axial\/} magnetic field profiles resemble 
those of the ILL 
experiment, with a few large spikes close to either end of the propagation region, their presence should not be a concern.

In the generalisation of these considerations to the \emph{full\/} magnetic field, one can parallel the discretisation method adopted (and tested) in 
Ref.~\cite{FlorianM.Piegsa2015} for the computation of the neutron spinor evolution operator in pulsed fields. Thus, the effect of a 
spike in the \emph{full\/} magnetic field 
on evolution is approximated (at the $k$th magnetic leak) as a rotation through an angle $\widetilde{\varphi}_k$ 
about a unit vector $\widehat{n}_k$. 
The associated unitary matrix is $ U_{2,k} = \exp\left( -\tfrac{i}{2}\widetilde{\varphi}_k\, \widehat{n}_k\cdot \vec{\sigma}\right)$.
The reasoning employed in the estimate of $\Delta\varphi_\text{max}$ above can be repeated to set a bound on the $|\widetilde{\varphi}_k|$'s, which, similarly,
indicates that they should be smaller for any of the experimental configurations considered in Ref.~\cite{Phillips2016} than in the ILL experiment.
    
\section{Inclusion of magnetic field in the vicinity of the detector} \label{sc:Dect}

A non-negligible magnetic field ($\gg 1\,\mu\text{T}$) is unavoidable in the space intervening between the end of the shielded quasi-free propagation 
region and the antineutron detector. In an analysis of the influence of this field, it is appropriate to distinguish between its spatial and temporal average 
$\vec{B}_\text{av}$ and fluctuations about $\vec{B}_\text{av}$. The uniform $\vec{B}_\text{av}$ can be discussed within the aid of 
Eq.~(\ref{eq:pvv1D}) by the formal device of aligning the $z$-axis with $\vec{B}_\text{av}$. The fluctuations are assumed to be small in relation to 
$\vec{B}_\text{av}$ and are ignored in the present analysis. [A non-perturbative analysis of fluctuations in the axial field based on Eq.~(\ref{eq:pvv1D})
shows that  there is little or no change in $\eta$ provided $B_\text{rms} \lesssim \tfrac{1}{2} B_\text{av}$.]

The effect of an uncompensated $\vec{B}_\text{av}$ can be readily gauged in a model in which the magnitude of the magnetic field has the idealised 
behaviour (in the rest frame) $B(t)= B_\text{av}\,\Theta\! \left( t- t_\text{qf} \right)$, where $t_\text{qf}$ is the time-of-flight for the quasi-free propagation 
region (of length $l_\text{qf}$). Then, beyond the quasi-free propagation region ($t>t_\text{qf}$), use of Eq.~(\ref{eq:pvv1D}) yields
\begin{equation}\label{eq:Ps}
 \frac{P_{\overline{n}}(t)}{\delta^2\, t_\text{qf}^2}  =  1 +  \frac{2}{\omega_{\text{av},L} t_\text{qf}} \sin\omega_{\text{av},L} (t - t_\text{qf} ) 
                                                                                     +  \frac{2}{(\omega_{\text{av},L} t_\text{qf})^2} \bigl[1-\cos\omega_{\text{av},L}(t - t_\text{qf}) \bigr]  ,
\end{equation}
where $\omega_{\text{av},L} = - \gamma B_\text{av}$.
Deviations from the quasi-free scaling term [which is the first term on the righthand-side of Eq.~(\ref{eq:Ps})] are negligible, being suppressed by inverse powers of 
\[
   |\omega_{\text{av},L}| t_\text{qf}  >  |\omega_{\text{av},L}| \frac{l_\text{qf}}{v_\text{min}} 
                                                       =  \alpha^{-1} |\omega_{\text{av},L}|\frac{l_\text{qf}}{l} \sim (10^2\,\text{$\mu$T}^{-1}) \, B_\text{av} ,
\]
where the last product evaluates to a thousand or more for the fields under consideration. In effect, $P_{\overline{n}}(t)$ is frozen at the value 
$P_{\overline{n}}(t_\text{qf})$ attained at the end of the interval of quasi-free propagation.

As a consequence of the stagnation in the value of $P_{\overline{n}}(t)$, the quasi-free propagation efficiency 
\begin{equation}\label{eq:qfvnqf}
  \eta = \eta_\text{qf} \frac{l_\text{qf}^2}{l^2}
\end{equation}
for $l > l_\text{qf}$, where $ \eta_\text{qf}$ is the value of this efficiency at the end of the quasi-free propagation region. The $l$-dependence of $\eta$ in 
Eq.~(\ref{eq:qfvnqf}) is stronger than that of $\eta$ in the quasi-free regime for the field strengths adopted in Fig.~\ref{fg:etaL}. 
For example, if $l_\text{qf}=175\,\text{m}$ (on average) and the distance of the detector from the end of the magnetically shielded region is $2.8\,\text{m}$ 
(as in the ILL experiment), then, at the location of the detector, $\eta= 0.97\eta_\text{qf}$. Over a distance of $2.8\,\text{m}$, $\eta$ has 
decreased by 3\%, whereas, over a quasi-free propagation distance of $175\,\text{m}$, $\eta_\text{qf}$ decreases by at most 2\% 
(see Fig.~\ref{fg:etaL}). Nevertheless, the quantitative difference between $\eta$ and $\eta_\text{qf}$ is sufficiently small that,
in any figure of merit, one can, in practice, substitute $\eta$ by $\eta_\text{qf}$, which is the quantity computed in Sec.~\ref{sc:qfpe}.

\section{Conclusion} \label{sc:Con}

The investigation of neutron-antineutron oscillations at the ILL in the early 1990's entailed construction of the largest high efficiency magnetically shielded system
in existence. In this paper, we have considered what the hard won experience at the ILL suggests about prospects for an even larger shielded system.
Concerning the effect of the large spikes in the residual field at the location of unavoidable magnetic leaks in the shielding system,
we have presented an argument, relying on little more than reasonable estimates of scaling with system size, that 
these features can be ignored because they could be ignored in the analy{\-}sis of the ILL experiment.
A calculation admitting magnetic fields of arbitrary strength implies that any large magnetic field encountered after the oscillating neutron-antineutron
system exits the quasi-free propagation region (specifically, the field surrounding the detector) will not degrade the probability for antineutron detection.
Instead, it is frozen at the value attained at the end of the propagation region.
Most of our attention, however, has been focussed on a perturbative treatment of the residual field (\emph{sans\/} spikes)
as a random process with the aim of clarifying the relation between the length $l$ of the propagation region and the quasi-free propagation 
efficiency $\eta$. Our findings establish that the dependence of $\eta$ on $l$ is approximately linear (cf.~Fig.~\ref{fg:etaL}).
The overall import of the related numerical estimates of $\eta$ for values of $l$ relevant to the design of future experiments is encouraging: 
to attain quasi-free propagation efficiencies in excess of $98\%$, 
it is enough to improve on the level of magnetic field suppression achieved in the ILL experiment by a mere factor of two.

Some aspects of our perturbative treatment of $\eta$ warrant further scrutiny. 
Variations in the residual magnetic field (after exclusion of any large spikes) have been assumed to be wide sense stationary. 
We believe that this assumption is justified for the system under discussion because of the compensation currents deployed, but, nonetheless, its
compatibility with data on magnetic field profiles should be tested.
Less fundamental to our analysis is the assumption that the magnetic field noise is Markovian, which serves to fix the frequency dependence of the required
power spectral densities [cf.~Eq.~(\ref{eq:mwssn})]. It would, of course, be better if power spectral densities taken from experiment
were employed, but our results on $\eta$ suggest that the precise functional form of these densities is unimportant for the estimates
made in this paper. More crucial is the ratio of the moments of the axial speed distribution in
our final expression for $\eta$ [in Eq.~(\ref{eq:etafinal})]. We have attempted to compensate for our ignorance about this ratio
by working with two radically different \emph{ansatzes\/} for the axial speed distribution [given in Eqs.~(\ref{eq:ns}) and (\ref{eq:nd})]. More 
precise estimates of $\eta$ require empirical constraints on this ratio of moments.
 
\begin{acknowledgments}
E.D.D. is very grateful to Prof.~Dr.~D.~Dubbers for supplying him with examples of magnetic field profiles measured during the ILL experiment, and
thanks the Institute of Nuclear Theory and the organizers of the INT program \emph{Intersections of BSM Phenomenology and QCD for New
Physics Searches\/} for their hospitality. This work was supported in part by the US Department of Energy under Grant No.~DE-FG02-97ER41042.
A.R.Y. also acknowledges support by the National Science Foundation (Grant No.~1307426).
\end{acknowledgments}

\end{document}